# A Self-Regulated and Reconfigurable CMOS Physically Unclonable Function Featuring Zero-Overhead Stabilization

Dai Li, *Student Member, IEEE*, Kaiyuan Yang, *Member, IEEE*

*Abstract*— This paper presents a reconfigurable physically unclonable function (PUF) design fabricated in 65nm CMOS technology. Subthreshold-inverter-based static PUF cell achieves 0.3% native bit error rate (BER) at 0.062 fJ/bit core energy efficiency. A flexible, native transistor-based voltage regulation scheme achieves low-overhead supply regulation with 6 mV/V line sensitivity, making the PUF resistant against voltage variations. Additionally, the PUF cell is designed to be reconfigurable with no area overhead, which enables stabilization without redundancy on chip. Thanks to the highly-stable and self-regulated PUF cell and the zero-overhead stabilization scheme, a 0.00182% native BER is obtained after reconfiguration. The proposed design shows 0.12%/10 °C and 0.057%/0.1 V bit error across military-grade temperature range from -55 °C to 125 °C and supply voltage variation from 0.7 V to 1.4 V. The total energy/bit is 15.3 fJ. Furthermore, the unstable bits can be detected by sweeping body bias instead of temperature during enrollment, significantly reducing the testing costs. Last but not least, the prototype exhibits almost ideal uniqueness and randomness, with a mean inter-die hamming distance (HD) of 0.4998 and a 1020× inter/intra-die HD separation. It also passes both NIST 800-22 and 800-90B randomness tests.

## I. INTRODUCTION

PHYSICALLY unclonable functions (PUFs) are increasingly studied and developed for secure electronic devices, especially for resource-constrained systems like Internet of Things (IoT) [1]-[3]. PUFs harvest intrinsic process variations of integrated circuits to generate keys and IDs unique to each device. It not only provides a low-cost solution to secure key generation and storage, but also offers attractive features such as bonding with specific hardware and tamper evidence, which makes them viable solutions for many emerging hardware security issues on supply chain tracking, counterfeit protection, system attestation, etc.

The generation and storage of secure keys are critical for entity authentication, secure communication, and serving as roots of trusts for computing systems. Ubiquitous IoT devices face additional challenges because of varying environmental conditions and physical access by attackers. An ideal solution should exhibit the following three properties. First, non-volatile storage of the keys under voltage and temperature variations is necessary. Second, low cost and energy consumption are essential for IoT devices with constrained resource and battery lifetime. Third, robustness and alertness against physical tampering attacks is highly desired.

Non-volatile memories (NVMs) are the conventional solutions for secret key storage. NVMs include one-time programmable read-only memories (ROMs), fuses, and programmable flash memories. While NVMs provide excellent reliability and long-term data storage, NVM-based key storage solutions suffer from the following drawbacks: (1) Most NVMs require extra fabrication steps, adding to higher fabrication costs; (2) Conventional ROM, fuses and flash memories are all vulnerable to physical attacks, such as optical imaging techniques and direct probing; (3) Software vulnerabilities can also be exploited to gain access to keys stored in NVMs with standard I/O interface.

Physically unclonable function (PUF) is the most promising alternative to conventional NVM and is expected to meet all the desired properties for secure key storage. Firstly, PUFs use intrinsic process variations to generate and store the keys. The keys are stored in device characteristics, rather than direct digital data storage, making it more difficult to directly read out stored keys with tampering attacks. Moreover, PUFs exploit small device variations, which are believed to be sensitive to physical tampering and make PUFs tamper-evident. Secondly, the keys are unique to each chip due to the random and chip-specific nature of process variations during chip manufacturing. Therefore, no key programming is required and cloning a known device is almost impossible, making PUF literally "unclonable". Thirdly, silicon PUFs are low cost and easy to integrate with modern system-on-chip (SoC), because they only require standard CMOS devices and are easily portable across process nodes. Lastly, PUFs can be designed to be highly area and energy efficient, making them suitable for systems ranging from IoT devices to high-performance SoCs.

Generally, PUFs are categorized into strong and weak ones based on their capabilities. Strong PUFs [4]-[8] provide a large number of challenge-response pairs (CRP) for direct authentication, but existing designs have limited stability and randomness, leading to vulnerabilities against machine learning attacks [9]. Recent works explored circuit nonlinearity to improve resiliency against machine learning attacks [6]-[7]. Weak PUFs have smaller capacity, but achieve higher stability against PVT variations and close-to-ideal uniqueness to provide reliable IDs and keys. This work focuses on weak PUFs and we do not differentiate strong and weak PUFs in the rest of the

2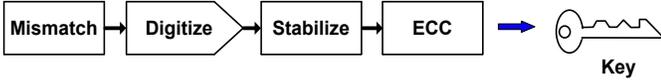

Fig. 1. Typical PUF diagram from entropy extraction to key generation.

paper. The following metrics are widely adopted to evaluate PUFs: *reliability, randomness, area, energy efficiency, and throughput*, out of which the stability over PVT variations is the major challenge to PUF designs.

As shown in Fig. 1, a PUF usually comprises of four stages. The random mismatch of transistors or other components, acting as the entropy source, is extracted and digitized into a binary data. The raw PUF responses then go through stabilization process and error-correcting code (ECC) to remove and correct all unstable bits, providing 100% reliable keys for security applications. In theory, all bit errors can be corrected by reserving enough redundant bits for stabilization and ECC under a given bit error rate (BER). However, the redundancy and complexity increase super-linearly with BER. According to [10], the energy requirement for ECC to correct one-bit error in a 256-bit key equals to the energy of accessing over 200 PUF cells. Therefore, achieving higher stability at early stages of the whole PUF processing flow is the key to improve the overall PUF performance, demanding better PUF cell design and stabilization method.

PUF design involves entropy extraction and digitization. Various circuit topologies have been proposed for entropy extraction, including metastable cross-coupled inverters [11], power-up state of SRAM cells [12], delay lines [3], oscillators [5], current mirrors [13], PTAT references [14], and leakage-based transistor pairs [15]. Comparing these PUF designs, circuits with static operations avoid random noise associated with dynamic transients, generally providing more reliable responses. In terms of digitization, comparators shared by one column or the whole array were commonly deployed for lower area overhead, but the comparator must be designed with high gain and accurate offset cancellation, which incurs high power and complexity. Even microvolt offsets lead to biased PUF responses and reduced uniqueness. Moreover, even with optimal offset cancellation, shared comparators suffer from long wires and coupling effects. Comparatively, if the digitizer is integrated into every cell locally [16], comparator offset becomes part of the entropy source and will not affect PUF uniqueness. Complementary current mirror-based monostable PUF design is one of the first implementations to combine static operation and local digitization [13] [17]. This fully static design provides strong resiliency to environmental variations and random noise because of the absence of dynamic switching events, at the expense of higher standby power and larger PUF cell footprint. Another local amplification scheme is proposed in [16], where a NAND gate chain amplifies threshold voltage differences between neighboring stages. This design obtained lowest BER among all reported CMOS PUFs and a compact footprint, but it consumes high short circuit power. In [18], a chain of sub-threshold 2-transistor amplifiers further improves the NAND chain design and achieves best-in-class metrics.

Numerous stabilization techniques have been proposed to correct or discard error bits, which usually comes with area and time overheads. Temporal majority voting (TMV) and spatial majority voting (SMV) filter out random noise by taking multiple bits in time or space domain for a single bit output. While the complexity for TMV and SMV is low, the enhancement of stability is limited. Burn-in through intentional aging [11] [19] improves stability with little area overhead but incurs high testing cost. Another widely used stabilizing technique is to find and filter out unstable PUF cells during enrollment. However, long testing time with temperature sweep are necessary to find all unstable cells. It also requires redundant cells in the PUF array on chip for replacement and on-chip or off-chip storage of masking map. Lastly, ECC such as BCH code [11] [16] is capable of 100% stability at the expense of high computing complexity, high redundancy and long latency.

To further improve PUF stability without extra area, power, and throughput overhead and sacrifice of technology portability, this paper presents a self-regulated and reconfigurable PUF design with zero-overhead stabilization scheme [20]. It features three major advantages:

- Subthreshold inverter-based static and local digitization structure exhibits ultra-low power consumption, state-of-the-art native stability, and compact footprint.
- Native transistor-based regulation provides subthreshold supply voltage for PUF cells with low overhead and further improves PUF's resistance to voltage variations.
- The proposed in-cell reconfiguration scheme enables 100 times BER reduction with no area overhead on chip.

The remainder of the paper is organized as follows. The proposed PUF circuit is presented in Section II. Section III explains modeling and implementation of the reconfiguration scheme for stabilization. In Section IV, measurement results are presented and discussed. Conclusions are drawn in Section V.

## II. PROPOSED PUF CELL DESIGN

In order to achieve high stability with low area and power overhead, the proposed PUF cell employs a 4-stage subthreshold inverter chain for entropy extraction and amplification, inspired by the static and local digitization designs in [16] and [18]. As shown in Fig. 2, the input and output of first stage is shorted, setting its voltage to a switching point with high gain. Mismatch between switching voltages of successive stages is amplified to full rail after a few stages. This PUF structure eliminates the impacts of non-ideal comparators and noise during dynamic transitions, leading to higher stability. However, the use of NAND gates at nominal VDDs incurs high power consumption and 2-transistor amplifiers are found to be less effective at more advanced technology nodes than 180nm. The following subsections will discuss the use of subthreshold inverters, low-overhead supply regulation of the PUF cell, and system implementation of the PUF module.

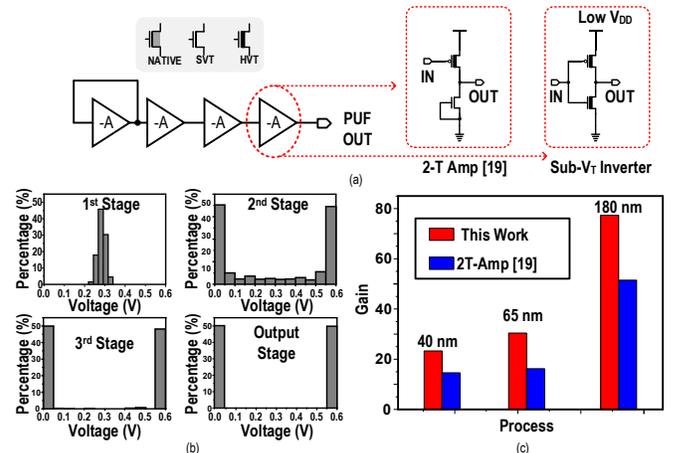

Fig. 2. (a) PUF cell topology, comparison of 2-transistor amplifier-based cell and proposed subthreshold inverter-based cell, (b) simulated histogram of voltages at each stage, (c) voltage gain of 2-transistor amplifier and subthreshold inverter in different technology nodes.

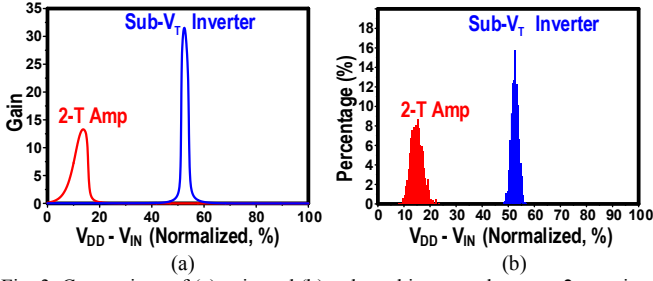

Fig. 3. Comparison of (a) gain and (b) voltage histogram between 2-transistor amplifier and subthreshold inverter.

*A. Subthreshold Inverter Based PUF Cell*

The stability of PUFs based on a chain of amplifiers depends on the distribution of switching voltage mismatch and the amplification gain of each stage. Firstly, higher gain improves overall PUF stability over environmental variations because less stages are involved in deciding the final responses. Secondly, larger mismatch variations apparently reduce the probability of having an unstable cell. Thirdly, if the gain around the switching voltage is asymmetrical, the PUF response will be biased towards one value and the overall randomness is affected. It is found that the 2-transistor amplifier in [19] shows reduced voltage gain at more advanced process nodes because of decreased output impedance. Because an off-transistor is used to bias the amplifier, the switching voltage is close to the supply voltage, leading to unbalanced gain and headroom.

As shown in Fig. 2, we propose the use of thick-oxide inverters under sub-threshold supply voltage, which provides higher gain and balanced output while consuming low static power. It is shown in Fig. 2(c) and Fig. 3 that as technology nodes scales, subthreshold inverter consistently yields higher voltage gain than 2-transistor amplifier, and shows symmetrical gain and headroom. Therefore, the proposed PUF cell achieves higher native stability, better randomness, and better portability to different process over the 2T-amplifier design in [19].

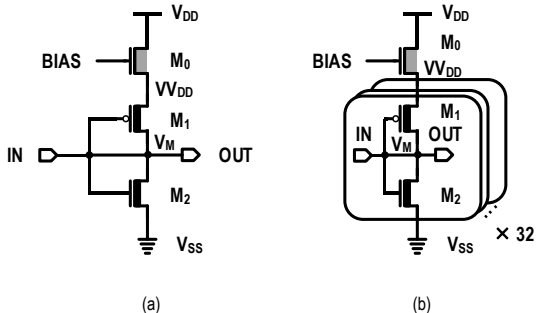

Fig. 4. (a) Cell-wise voltage regulation model using a native transistor, (b) column-wise voltage regulation model using a native transistor.

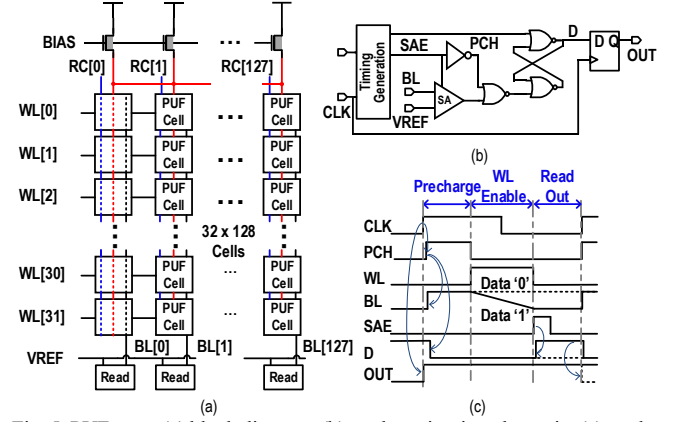

Fig. 5. PUF array (a) block diagram, (b) readout circuits schematic, (c) readout timing.

*B. Voltage Regulation using Native Transistors*

Stable and efficient low voltage supply is required for the subthreshold inverter-based PUF cell for stability and low energy consumption. One of the main sources of instability for PUF comes from voltage variation, especially in subthreshold PUF designs [8]. IR drop, EM interference and other incidents can cause fluctuation of supply voltage in PUF cells.

A conventional approach to provide a regulated low supply is to use a low drop-out regulator (LDO). LDOs have quiescent currents, leading to low efficiency and large area overhead when the load is light. PUFs can be offered in either a large array or as small embedded key registers. Thus, a more scalable and low-overhead supply regulation is desired for our proposed PUF circuits.

Inspired by a threshold-based voltage reference [21], this work utilizes a native transistor to self-regulate supply voltage with low area and power overhead. The voltage regulation works on the basis that the cell is biased in subthreshold region. The first stage consumes most of the current and the rest stages are almost in cutoff region. Therefore only the first stage is considered in the following calculations. The equivalent model is shown in Fig. 4. The subthreshold region currents of $M_0$ and $M_2$ are derived in (1) and (2), according to [21], where $I_{sub}$ denotes the cell current flowing through $M_0$, $M_1$ and $M_2$ and should be equal in (1) and (2). $V_M$ denotes the switching voltage of stage 1. $VV_{DD}$ denotes the regulated virtual supply voltage of the inverters.

$$I_{sub} = \mu_0 C_{OX0} \frac{W_0}{L_0}(m_0 - 1)V_T^2 \exp\left(\frac{V_{BIAS} - VV_{DD} - V_{th0}}{m_0 V_T}\right)\left(1 - \exp\left(\frac{-V_{ds0}}{V_T}\right)\right) \quad (1)$$

$$I_{sub} = \mu_2 C_{OX2} \frac{W_2}{L_2}(m_2 - 1)V_T^2 \exp\left(\frac{V_M - V_{th2}}{m_2 V_T}\right)\left(1 - \exp\left(\frac{-V_{ds2}}{V_T}\right)\right) \quad (2)$$

$V_{DS}$ in (1) and (2) are by magnitude larger than $V_T$ thus the last terms can be neglected. To simplify the model, it is assumed that $V_M$ is half of $VV_{DD}$. In real design, $V_M$ does not necessarily need to be exactly half of $VV_{DD}$.

$$V_M = \frac{1}{2} VV_{DD} \quad (3)$$

The equation for VV$_{DD}$ can be derived in (4), showing independence of supply voltage. Moreover, since thermal voltage is proportional to temperature and threshold voltage is complementary to temperature. The temperature effects of the transistors on VV$_{DD}$ can be cancelled by carefully sizing the native transistor.

$$VV_{DD} = \frac{2m_0 m_2}{m_0 + 2m_2} V_T \ln\left(\frac{\mu_0 C_{OX0} W_0 L_2 (m_0-1)}{\mu_2 C_{OX2} W_2 L_0 (m_2-1)}\right) + \frac{2m_0}{m_0+2m_2} V_{th2} + \frac{2m_2}{m_0+2m_2}(V_{BIAS} - V_{th0}) \quad (4)$$

Further area saving is achieved by column-wise sharing of the native regulation transistor in this design. The flexible configuration of the native regulation transistor can adapt to cell-wise, column-wise and array-wise regulation. In the proposed design, each column of 32 cells share a native regulating transistor. The VV$_{DD}$ equation for this case is modified as in (5).

$$VV_{DD} = \frac{2m_0 m_2}{m_0 + 2m_2} V_T \ln\left(\frac{\mu_0 C_{OX0} W_0 L_2 (m_0-1)}{32\mu_2 C_{OX2} W_2 L_0 (m_2-1)}\right) + \frac{2m_0}{m_0+2m_2} V_{th2} + \frac{2m_2}{m_0+2m_2}(V_{BIAS} - V_{th0}) \quad (5)$$

### C. System Implementation

The block diagram of this design is depicted in Fig. 5(a). A 32 by 128 array is implemented. SRAM-style peripherals are integrated to allow for parallel and high-speed readout. Each column shares the same virtual supply voltage regulated by a native transistor. Readout process involves BL pre-charging, WL enabling, voltage sensing through a single-ended sense amplifier (SA), and latching of the result. A waveform for reading operations is included in Fig. 5(c).

## III. ZERO-OVERHEAD RECONFIGURATION SCHEME

Conventional stabilization methods, as discussed before, could not achieve good balance between overhead and stabilization effect. A zero-overhead reconfiguration scheme is proposed to stabilize the PUF cell. The scheme is based on the specific structure of the proposed PUF cell, and can be enrolled with low testing cost. In this work, the V$_M$ difference between the first two stages is the entropy source. Cell-level reconfiguration involves transistor merging and converting the original cell into a new cell with independent performance.

### A. Source of Instability

The entropy source of PUFs is the local mismatch due to process uncertainty. The randomness in dopant distribution, lithography, gate thickness causes variations of transistors. Ideally the switching voltages of two inverters within the same PUF cell should be fixed, leading to a stable '1' or '0' output. However, the actual mismatch can be altered or even flipped under voltage and temperature variations, especially when the original mismatch is small. This is caused by the fact that different transistors exhibit varying environmental sensitivities.

TABLE I
PROBABILISTIC MODEL FOR RECONFIGURABLE PUF CELL

| Cell Mismatch | Probability | Action | Output |
|---|---|---|---|
| V$_{M1}$ > V$_{M2}$ | 0.5*(1-P$_O$) | Stay | '0' |
| V$_{M1}$ < V$_{M2}$ | 0.5*(1-P$_O$) | Stay | '1' |
| V$_{M1}$ ≈ V$_{M2}$ > V$_{M3}$ | 0.5*P$_O$*(1-P$_R$) | Reconfigure | '1' |
| V$_{M1}$ ≈ V$_{M2}$ < V$_{M3}$ | 0.5*P$_O$*(1-P$_R$) | Reconfigure | '0' |
| V$_{M1}$ ≈ V$_{M2}$ ≈ V$_{M3}$ | P$_O$*P$_R$ | Reconfigure | Unstable |

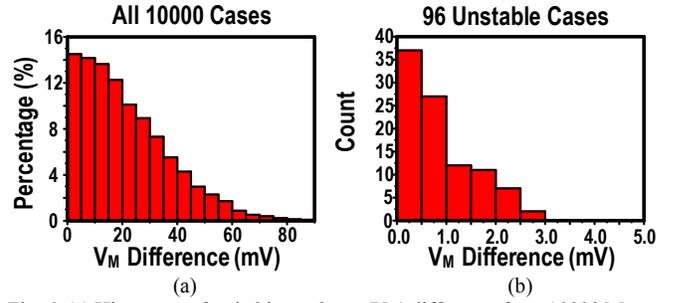

Fig. 6. (a) Histogram of switching voltage (V$_M$) difference from 10000 Monte-Carlos simulations, (b) from 96 unstable cases.

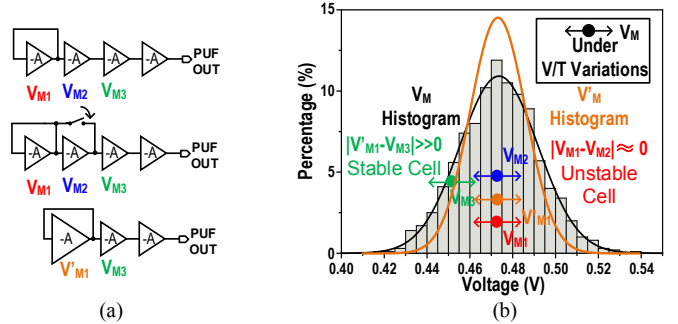

Fig. 7. (a) Process of in-cell reconfiguration of an unstable cell, (b) the switching voltages of an originally unstable cell before and after reconfiguration.

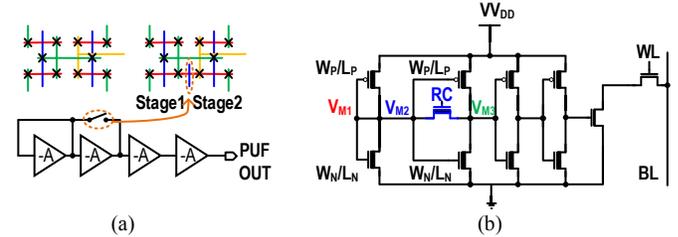

Fig. 8. Reconfigurable PUF cell in (a) layout, (b) schematic views.

Monte-Carlo simulations of 10,000 proposed PUF cells are performed to investigate the cause of instability. More focus is put on temperature variations because the proposed native voltage regulation suppresses the impacts of voltage variation. Simulation results show that 96 out of 10,000 cells have unstable output under -55 °C to 125 °C temperature sweep. The histogram of switching voltage differences of stage1 and stage2 illustrate that all the 96 unstable cases have switching voltage differences smaller than 3 mV, as shown in Fig. 6. Conclusion can be drawn from above discussions that small mismatch is a necessary condition for an unstable PUF cell.

### B. In-Cell Reconfiguration

The reconfiguration process is shown in Fig. 7 to further illustrate the source of instability and explain the proposed stabilization scheme

An originally unstable cell has small mismatch between the first and second stage. The switching voltages of the first three stages of this cell are denoted by V$_{M1}$, V$_{M2}$ and V$_{M3}$ respectively. V$_{M1}$ and V$_{M2}$ are close to each other in nominal condition and are likely to flip their relative difference under V/T variations. Based on the observation that local mismatch is purely random, the reconfiguration scheme is proposed by combining the first two stages as a new first stage. The output of the reconfigured cell depends on the difference of the combined new stage and the third stage. The subthreshold inverter-based cell has a low bit error rate. Thus, the probability of $V_{M1}$ and $V_{M2}$ being close is low. For the reconfigured cell, the necessary condition of it being unstable over V/T variations is that $V_{M1}$, $V_{M2}$ and $V_{M3}$ are all close to each other. The probability is extremely low in the proposed design.

Table I describes the probabilistic model of the reconfiguration scheme. $P_O$ and $A_{VO}$ denotes the probability of unstable bits and gain for original cell. $P_R$ and $A_{VR}$ denotes the probability of unstable bits and gain for reconfigured cell. $V_M$ and $\sigma$ is the mean and standard deviation of a subthreshold inverter switching voltage The bit error rate and voltage gain have a relationship of $f(A_v)$. The probabilistic model of the process in shown in Table I. The switching voltages of the original cell and reconfigured cell exhibit the following distribution. It is assumed that transistor threshold voltage $V_M$ is independent and identically distributed.

$$V_{M1}, V_{M2}, V_{M3} \sim N(V_M, \sigma) \quad (6)$$

According to [22], the standard deviation of threshold voltage is inversely proportional to transistor area, which leads to equation (7).

$$V'_{M1} \sim N(V_M^*, 0.707\sigma) \quad (7)$$

$V'_{M1}$ denotes the switching voltage of stage1 in the reconfigured cell. $V_M^*$ denotes the mean of a combined subthreshold inverter's switching voltage. The difference of switching threshold voltage between the first two stages can be derived in (8) and (9) for the original and reconfigured cell.

$$V_{M1} - V_{M2} \sim N(0, 1.414 * \sigma) \quad (8)$$
$$V'_{M1} - V_{M3} \sim N(V_M^* - V_M, 1.224 * \sigma) \quad (9)$$

The probability of unstable bits is inversely proportional to the standard deviation of first two stages' switching threshold voltage difference.

$$P_O \propto \frac{f(A_{vO})}{[\mathrm{Var}(V_{M1} - V_{M2})]^{\frac{1}{2}}} \quad (10)$$

$$P_R \propto \frac{f(A_{vR})}{[\mathrm{Var}(V'_{M1} - V_{M3})]^{\frac{1}{2}}} \quad (11)$$

From (10) to (11), $P_R$ is estimated to be $1.15*f(A_{VR})/f(A_{VO})*P_O$. In measurements, the portion of cells requiring reconfiguration is 4% under -55 °C to 125 °C temperature variation and 0.7 V to 1.4 V supply voltage variation. The reconfigured cell has slightly higher BER than the original cell due to decreased mismatch induced by doubled size of the first stage [22]. Reduced gain due to reduced number of stages also induces higher instability. However, the overall probability for a cell to be unstable is the product of the two probability in theory, which is 0.44% in measurement.

TABLE II
COMPARISON OF STABILIZATION METHODS

| Stabilization Method | Testing Cost | Redundancy | Runtime Latency | Efficacy |
|---|---|---|---|---|
| Temporal Majority Voting | Low | Low | Medium | Low |
| Burn-in | High | Low | Low | Medium |
| Mask (Filter) | High | Medium | Low | High |
| Error-Correcting Code | Low | High | High | Highest |
| Proposed Reconfiguration | Medium | Low | Low | High |

### C. Physical Implementation of Reconfiguration Scheme

The benefit of the proposed reconfiguration scheme is its low cost in this specific design. To combine the first and the second stage, an NMOS transistor is inserted between the drains of the two NMOS. Fig. 8 depicts the physical insertion of the reconfiguration transistor. In 65nm CMOS process, no area overhead is added under the design rules. In most technologies, the area overhead is also expected to be small. The reconfigure transistor is driven by full swing signal to connect two stages without voltage bias induced by $V_{TH}$ drop. As shown in Fig. 5, reconfiguration control signals are applied vertically. For each PUF array, a reconfiguration map (R-MAP) is maintained in the server. And upon fetching a specific key, its corresponding reconfiguration setting is applied to the array. No on-chip redundancy and low runtime latency is added in this scheme, in contrast to masking and ECC approach.

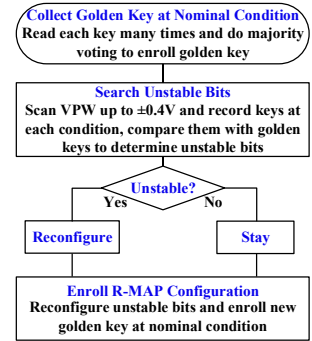

Fig. 9. Search and enrollment process of R-MAP configuration.

### D. Fast R-Map Searching via Body Bias Emulation

Ideally the R-MAP should be obtained by sweeping temperature and voltage in the desired operation range during enrollment. However, this approach consumes considerable testing cost and is expensive for massive production.

$$V_{th} = V_{th0} + \gamma \left( \sqrt{|2\phi_F + V_{SB}|} - \sqrt{|2\phi_F|} \right) \quad (12)$$

$$\phi_F = \frac{kT}{q} \ln\left(\frac{N_A}{N_i}\right) \quad (13)$$

A key observation from (12) and (13) for the proposed PUF is that its entropy source, threshold voltage, can be modulated by adjusting body bias to emulate temperature sweep. By properly sweeping body bias, the effects of temperature variation can be emulated and PUF cells with small mismatch may flip their outputs. By denoting such cells for reconfiguration, the necessary condition of them being unstable is eliminated, increasing the overall stability for PUF key generation. The R-Map search and enrollment process is shown in Fig. 9. In this design, deep n-well is used to isolate PUF array from external noise and enable PMOS body bias sweep. The amortized area overhead is negligible for each PUF cell. Emulating temperature variation using body bias (EVB) is fast and works at nominal condition, eliminating the need for expensive and time-consuming temperature sweep.

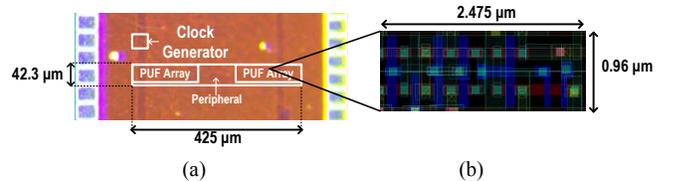

Fig. 10. (a) PUF chip micrograph, (b) cell layout.

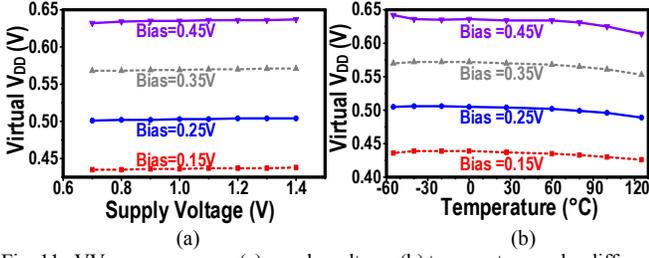

Fig. 11. VV$_{DD}$ curve versus (a) supply voltage, (b) temperature under different bias voltages.

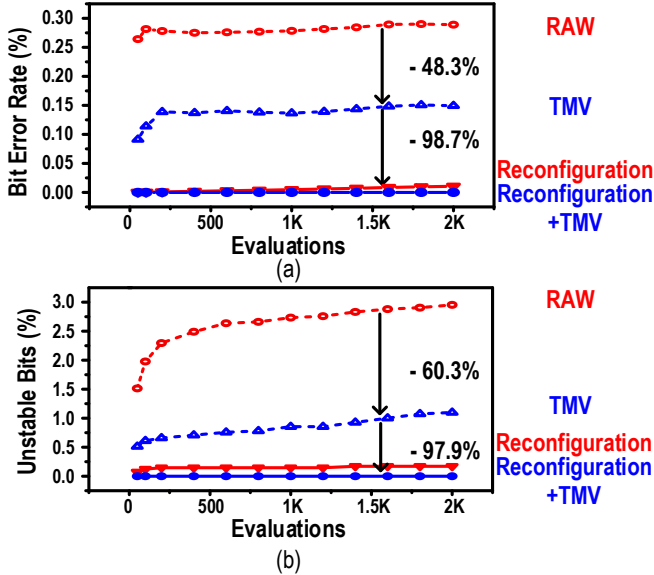

Fig. 12. (a) Bit error rate, (b) percentage of unstable bits versus number of evaluations under nominal condition.

The comparison of the proposed reconfiguration scheme and conventional methods is listed in Table II. It is highly effective with no on-chip area overhead, while adding low runtime latency and moderate test cost with the help of body bias.

## IV. MEASUREMENT RESULTS

The chip is fabricated in 65nm CMOS process. The 32 by 128 cells array occupies 0.018 mm$^2$. The die micrograph and the layout of the PUF cell are shown in Fig. 10. Each PUF cell measures 0.96 μm by 2.475 μm, or 562 F$^2$. Clock generator is integrated to provide high-speed clocking for key access.

The nominal condition for the PUF chip is 27 °C and 1.2 V supply voltage. Golden keys are collected at nominal condition by averaging our random noise with many samples. BER and unstable bit percentage results are measured by comparing separately collected samples under nominal and V/T variations with the golden key.

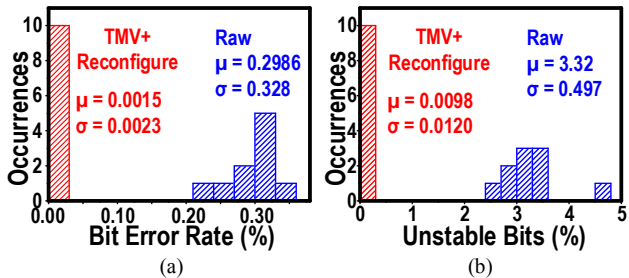

Fig. 13. (a) Native BER, (b) unstable bits histogram of 10 chips.

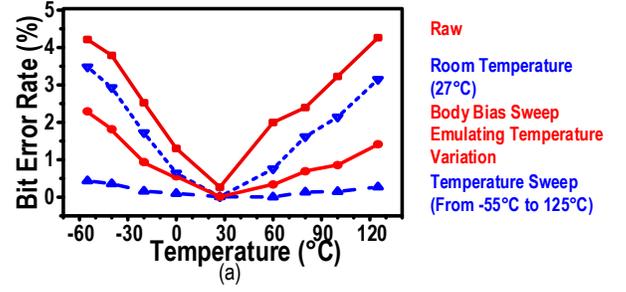

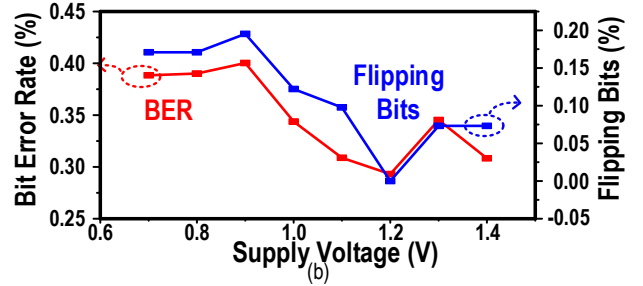

Fig. 14. (a) Bit error rate versus temperature variation under different stabilization methods, (b) BER and flipping bits versus supply voltage.

### A. Voltage Regulation

The reliability of the proposed native regulation is essential to the quality of the PUF. VV$_{DD}$ is measured across supply voltage and temperature sweep to evaluate the efficacy of native regulation. The measured line sensitivity of VV$_{DD}$ is less than 6mV/V over supply voltage range of 0.7 V to 1.4 V. Additionally, VV$_{DD}$ shows less than 10mV variation across -55 to 125 °C. For testing purpose, bias voltage is provided through off-chip source. For system-level applications, it can be generated by a 2-transistor voltage reference [21] which is also robust against voltage and temperature variations. No extra current is consumed for voltage regulation with the use of native transistor regulation.

### B. Native PUF Stability

The bit error rate (BER) and the proportion of unstable bits of 5 chips, measured at nominal condition before and after stabilization, are depicted in Fig. 12. BER denotes rate of bit flips over evaluations compared with golden value and unstable bit denotes percentage of ever-flipped bits. The definitions are adopted from [15]-[18] for fair comparison. It should be noted that the proportion of unstable bits increases as the number of evaluations increases re shown for fair. The native BER reaches steady state quickly after number of evaluations increases. The raw native BER is 0.30% without any stabilization methods. Temporal majority voting (TMV) reduced BER by 48.3% with 11 samples for 1 output. The reconfiguration methods followed by TMV can further reduce the BER to 0.00182%. The percentage of unstable bits is 2.95% after 2000 evaluations before stabilization. TMV11 alone reduces the percentage by 60.3% percent. Reconfiguration followed by TMV11 reduce the native unstable bits to 0.024%. More than 100 times BER improvement is achieved with low-overhead TMV11 and in-cell reconfiguration schemes.

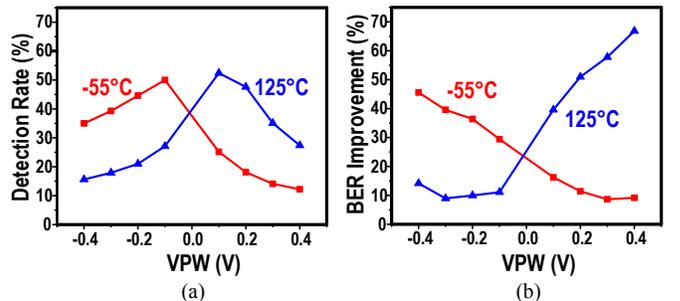

Fig. 15. (a) Detection rate, (b) BER improvement versus body bias sweep.

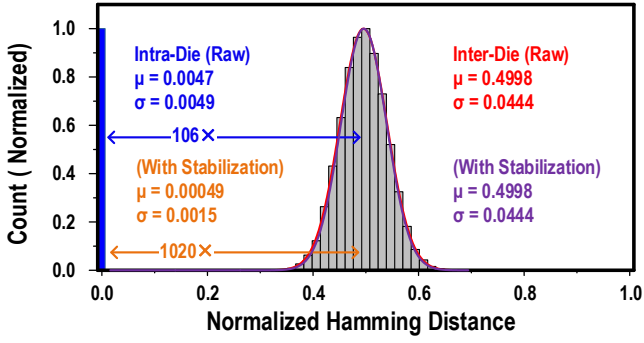

Fig. 16. Normalized Hamming Distance before and after reconfiguration.

### C. PUF Stability over Voltage and Temperature Variations

The stability of the proposed PUF is evaluated under the military-grade temperature range from -55 °C to 125 °C. The bit error rate (BER) over temperature sweep is 4.26% before stabilization. TMV and reconfiguration are applied to improve stability. Fig. 14(a) presents BER across temperature variation when different stabilization methods are applied. By detecting unstable bits under room temperature, a portion of bits that will be unstable over temperature variation is filtered out. However, this only achieves 25% BER reduction. Since unstable cells under temperature variation are not necessarily natively unstable cells. Filtering out all unstable cells by sweeping temperature during enrollment achieves the highest stability, reducing BER to 0.44%. The remaining errors are caused by random noise and cannot be stabilized, the stability is still among best in class. For stability-first applications, higher stability can be obtained at the cost of higher testing cost.

The proposed body-bias sweeping approach, named EVB, optimizes the tradeoff between stabilization effect and cost. BER is reduced to 2.27% by sweeping p-well biasing voltages (VPW) from -0.4 V to 0.4 V searching for unstable bits for reconfiguration. The process does not involve temperature sweeping thus has much lower testing cost than searching the full temperature range.

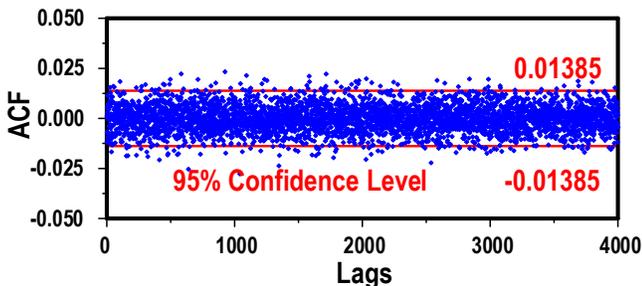

Fig. 17. Autocorrelation of 40960 bits for up to 4000 lags.

TABLE III
NIST PUB 800-90B RESULTS

| NIST Pub 800-90B (Draft 2) | Results of 10 chip × 4096 bits |
|---|---|
| IID Permutation | PASS |
| Chi-square Independence | PASS (score=1821.68, dof=2047) |
| Chi-square Goodness-of-fit | PASS (score=6.904, dof=9) |
| LRS Test | PASS (Pr=0.792) |

TABLE IV
NIST PUB 800-22 RESULTS

| NIST Pub 800-22 (rev. 1a, 2010) | $X^2$ of p-value | Average p-value | Pass Rate |
|---|---|---|---|
| Frequency | 0.097 | 0.456 | 96.67% |
| Block Frequency | 0.469 | 0.454 | 98.67% |
| Cumulative Sum-1 | 0.271 | 0.493 | 97.30% |
| Cumulative Sum-2 | 0.271 | 0.488 | 98.67% |
| Runs | 0.350 | 0.499 | 98.67% |
| Longest Runs | 0.686 | 0.509 | 100.00% |
| FFT | 0.437 | 0.462 | 98.67% |
| Serial-1 | 0.630 | 0.497 | 96.00% |
| Serial-2 | 0.779 | 0.478 | 100.00% |
| Approximate Entropy | 0.469 | 0.458 | 99.30% |
| Non Overlapping Template | PASS | PASS | PASS |

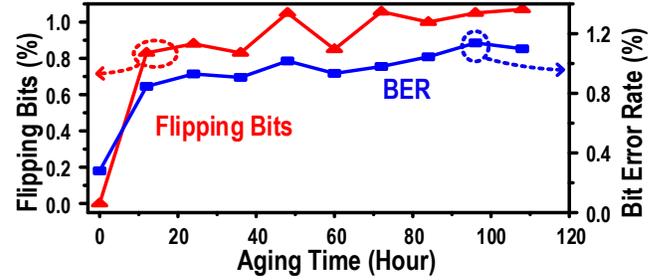

Fig. 18. Aging effects on BER and percentage of flipping bits.

Fig. 15 shows the detection rate and BER improvement of EVB at different VPWs. The detection rate is defined as the ratio of number of truly unstable bits across temperature variation filtered by EVB, over the number of total filtered bits by EVB at one VPW. Higher detection rate is observed for low VPW. This is expected because low VPW filters out cells with small mismatch, which is more likely to flip when temperature changes. Although detection rate lowers as body bias voltage rises, the overall stability still improves as shown in Fig. 15(b). High VPW filters out additional cells with larger mismatch, which will still flip under temperature variation, but with a lower probability. The BER improvement curve shows that positive VPW is more effective for high temperature while negative VPW is more effective for low temperature. This proves the assumption that the main source of entropy is mismatch of threshold voltage. Since positive VPW and high temperature both decrease threshold voltage, and negative VPW and low temperature both increase threshold voltage.

In addition to temperature variation, we also evaluate the PUF's resistance to supply voltage variation from 0.7 V to 1.4 V. Only 0.4% BER is observed across the voltage range, thanks to the supply regulation based on native transistors.

TABLE V
COMPARISON WITH PRIOR ARTS

| | THIS WORK (27°C 1.2V) | JSSC' 18 [17] | ISSCC' 18 [15] | JSSC' 17 [19] | ISSCC' 17 [18] | JSSC'16 [14] | ISSCC' 16 [16] | ISSCC'14 [11] |
|---|---|---|---|---|---|---|---|---|
| Technology | 65nm | 40nm | 180nm | 14nm | 180nm | 65nm | 45nm | 22nm |
| PUF Cell Area/Bit ($F^2$) | 562 | 3643 | 890 | 9387 | 782 | 726 | 2613 | 9628 |
| Native Unstable Bits (Evaluations) | 2.95% (2K) | 2.55% (500) | 5.62% (1K) | ~27%[f] (5K) | 1.73% (2K) | 6.54% (500) | - | 30% (5K) |
| Native Bit Error Rates | 0.30% | 0.81% | 0.69% | 5.76%[f] | 0.18% | - | 0.1%[b] | 8.3% |
| Stabilization Method | TMV, EVB, Reconfigure | Hysteresis, T Compensation | Remapping, TMV | Burn-in, TMV, SBD[c] | Masking, TMV | TMV, Masking | Valid Map, SMV[d], ECC | Burn-in, TMV, Masking |
| Unstable Bits After Stabilization[a] | 0.024% | - | 0.08% | - | 0.69% | 2% | - | 4.6% |
| BER after Stabilization[a] | 0.00182% | - | 0.019% | 1.46%[f] | 0.08% | - | - | 0.97% |
| Tested Conditions — Temp (°C) | -55~125 | -40~125 | 0~80 | 25~110 | -40~120 | 0~80 | -25~85 | 25~50 |
| Tested Conditions — Supply (V) | 0.7~1.4 | 0.8~1.0 | 1.2~1.8 | 0.55~0.75 | 0.8~1.8 | 0.6~1.2 | - | 0.7~0.9 |
| Bit Errors per 10°C | 0.12%[e] | 0.32% | - | ~0.17% | 0.21% | 0.44% | 0.15% | - |
| Bit Errors per 0.1V | 0.057%[e] | 0.72% | - | - | 0.29% | 0.13% | - | 0.49% |
| Fractional Inter-PUF HD | 0.4998 | 0.4907 | 0.50001 | 0.486 | 0.499 | 0.5001 | 0.498 | 0.49 |
| Entropy | 0.9998 | 0.9972 | - | 0.99993 | - | 0.9998 | 0.999998 | 0.9997 |
| Bit Rate (Mb/s) | 8592 / 1459[h] | 24000 | 0.018 | - | 4832 | 10.2 | 1.92 | 2000 |
| Core Energy (fJ/bit) | 0.076 / 0.062 | 1.02 | - | 4 | 13.5 | - | - | 13 |
| Total Energy[g] (fJ/bit) | 25.6 / 15.3 | 56.5 | 3600 | - | 91.1 | 548 | - | - |

a. At nominal condition, before using redundancy-based stabilization methods such as masking
b. Using glitch detection
c. Selective Bit Destabilization
d. Spatial Majority Voting
e. Results after EVB
f. Under V/T Variations
g. Including readout, timing and WL driver power
h. HP Virtual $V_{DD}$ ~ 0.57V LP Virtual $V_{DD}$ ~ 0.44V

*D. Uniqueness and Randomness*

The inter-die and intra-die hamming distances are depicted in Fig. 16. The inter-die hamming distance measured over 10 chips has mean value of 0.4998 which is near-ideal. The mean intra-die hamming distance before stabilization is 0.0047, achieving 106 times separation between inter- and intra-die hamming distances. After stabilization (reconfiguration followed by TMV), the intra-die hamming distance is 0.00049, showing state-of-the-art identifiability. The autocorrelation of 40960 PUF bits with the 95% white noise confidence level at 0.01385 is shown in Fig. 17. The near-ideal hamming distance and autocorrelation results validate the uniqueness of the proposed PUF.

To further evaluate the randomness of PUF responses, NIST 800-90B [23] and 800-22 [24] randomness test suites are performed on 40,960 bits collected from 10 chips. With the limited number of bits, 10 out of 15 subtests in 800-22 are available. NIST recommended settings were used to run the tests. More detailed definition and explanations of the testing parameters can be found in [23] [24]. The PUF bits passed all available sub-tests in the two suites, showing high-quality randomness as shown in Table III and Table IV.

*E. Aging Effects*

Another factor to consider for PUF stability is aging. Aging degrades PUF stability or completely flip bits. The main sources of aging effects in PUF are NBTI and HCI [14] [17] [25] [26]. In order to evaluate aging impacts, accelerated aging is applied by stressing the PUF at 150 °C and 1.4 V supply voltage. Every 12 hours, measurement was taken at nominal condition. A stressing of 108 hours in total was applied, resulting in equivalent effects as several years' aging under nominal condition. Since this design works in subthreshold region, it is expected to suffer minor aging effects. The measured aging-induced instability shown in Fig. 18 is similar to previously best reported results [14] [17].

*F. Throughput and Energy Efficiency*

The design reaches 8.592 Mb/s readout throughput in high performance (HP) mode and 1.459 Mb/s in low power (LP) mode. This is enabled by SRAM-style array and readout peripheral. The subthreshold operation consumes 0.076 fJ/bit core energy. The total energy including that of core and peripheral circuits is 25.6 fJ/bit in HP mode and 15.3 fJ/bit in LP mode. The throughput and energy efficiency curves versus $VV_{DD}$ are depicted in Fig. 19. The throughput and energy efficiency were measured at nominal condition. Since the circuits work in subthreshold region, transistor current increases with temperature exponentially. Native transistor-based regulation suppresses supply voltage-induced influence on current.

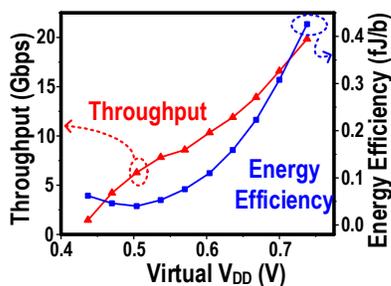

Fig. 19. Throughput and energy efficiency versus $VV_{DD}$ in nominal condition.

*G. Related Works and Comparison Table*

As shown in previous sections and Table V, the proposed PUF shows state-of-the-art native stability, area, and energy efficiency. The zero-overhead stabilization method further improves the stability of the PUF without the use of redundancy-based ideal masking and ECC.

Emerging NVM-based PUFs using anti-fuses and RRAMs [27] [28] provides almost zero BER by randomly generating keys and storing them in memories. This class of PUF does not preserve the sensitivity to tampering as in CMOS PUFs and usually require extra fabrication steps and high testing costs. Thus, these designs are not included in the comparison table but they represent a new promising direction in PUF design.

V. CONCLUSION

In conclusion, this paper presents a self-regulated PUF based on a subthreshold inverter chain, achieving 0.3% native BER and 0.076 fJ/bit energy efficiency. Native regulation provides resistance to supply voltage variation and lead to 0.057 %/0.1 V BER sensitivity against voltage variations. The proposed in-cell reconfiguration scheme reduced native BER by two orders of magnitudes to 0.00182% with no area overhead. Moreover, a fast searching method of emulating temperature variation by sweeping body bias is applied to locate unstable bits with low testing cost. The PUF prototype in 65nm occupies only 562 $F^2$ per bit. Measured responses from 10 chips pass all applicable NIST 800-22 and 800-90B randomness tests and show 1020 times separation of intra/inter-die Hamming Distances after stabilization. The design achieves best-in-class metrics in all desired properties, making it suitable to provide low-cost, high-performance key generation and storage for a wide range of applications.

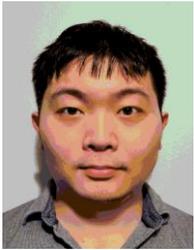

**Dai Li** received his BS and MS degree in Electronic Engineering from Tsinghua University and MS of Electrical and Computer Engineering from Rice University in 2010, 2013 and 2017 respectively. He is currently pursuing his PhD degree in Rice University.

His research interests include VLSI circuits, hardware security, mixed-signal integrated circuits and low-power circuits.

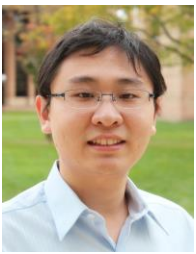

**Kaiyuan Yang** (S'13-M'17) received the B.S. degree in Electronic Engineering from Tsinghua University, Beijing, China, in 2012, and the Ph.D. degree in Electrical Engineering from the University of Michigan, Ann Arbor, MI, in 2017. His Ph.D. research was recognized with the 2016-2017 IEEE Solid-State Circuits Society (SSCS) Predoctoral Achievement Award.

He is an Assistant Professor of Electrical and Computer Engineering at Rice University, Houston, TX. His research interests include digital and mixed-signal circuits for secure and low-power systems, hardware security, and circuit/system design with emerging devices. Dr. Yang received the Distinguished Paper Award at the 2016 IEEE International Symposium on Security and Privacy (Oakland), the Best Student Paper Award (1st place) at the 2015 IEEE International Symposium on Circuits and Systems (ISCAS), the Best Student Paper Award Finalist at the 2019 IEEE Custom Integrated Circuits Conference (CICC), and the 2016 Pwnie Most Innovative Research Award Finalist.